\documentclass{aa}
\usepackage{natbib}
\usepackage{graphicx}
\bibpunct{(}{)}{;}{a}{}{,}

\begin{document}
\title{Searching for unknown open clusters in the Tycho-2 catalog
  \thanks {Based on observations of the ESA Hipparcos satellite} 
}

\author{
 B. S. Alessi\inst{1} 
\and 
A. Moitinho\inst{2,3}
\and 
W. S. Dias\inst{1,3} 
}

\offprints{W.S. Dias; \email{wilton@usp.br}}
\institute{
Universidade de S\~ao Paulo, Dept. de Astronomia,  CP 3386, 
S\~ao Paulo 01060-970, Brazil
\and Observatorio Astron\'omico Nacional, UNAM,
Apdo. Postal 877, C.P. 22800, Ensenada B.C., M\'exico
\and
CAAUL, Observat\'orio Astron\'omico de Lisboa, Tapada da Ajuda,
  1349-018 Lisboa, Portugal
}

\date{Received <May 23, 2003>/ Accepted <date>}

\abstract{We present 11 new open cluster candidates found in a
  systematic search for unknown star clusters using the astrometric
  and photometric data included in the Tycho2 catalog. The possible
  existence of these stellar aggregates is supported by the analysis
  of proper motions, color-magnitude diagrams, stellar density
  distributions, and by the visual inspection of the Digitized Sky
  Survey (DSS) plates. With these tools we were able to determine mean
  absolute proper motions as well as preliminary reddenings, distances
  and ages for the majority of the candidates.  We found that most of
  them are possibly nearby (closer than $\sim$600 pc) open clusters
  never studied before.   }

\titlerunning{New open cluster candidates}
\authorrunning{Alessi B. S. et al.}
\maketitle

\section{Introduction \label{sec:intro}}
The new generation of high precision astrometric catalogs such as
Hipparcos \citep{ESA1997} and Tycho2 \citep{Hog2000} offers an
important opportunity to study the kinematics of the nearby known open
clusters and associations
\citep{Dias2001,Baumgardt2000,Robichon1999,deZeeuw1999}. Moreover,
these catalogs also provide data for systematic searches for new
open cluster candidates based on proper motions and trigonometric
parallaxes \citep{Platais1998}.

In this work, we present the first results of a search for previously
unknown open clusters in the solar neighborhood.  The search is based
on the analysis of observational data (proper motions, photometry and
images) provided by the Tycho2 catalog, and by the Digitized Sky
Survey (DSS) plates.

The cluster candidates described here are all previously unknown, or
at least unlisted, according to the the most up-to-date catalog of
open clusters presently available \citep{Dias2002cat}.  The objects
were named Alessi~1, 2, 3, etc, sorted by right ascension.  These
clusters are all large (20$^{\prime}$ - 110$^{\prime}$), rather bright
and poor, indicating that they are possibly nearby sparse clusters or
some sort of Open Cluster Remnants (hereafter OCR). As we will show
later on, most candidates are possible nearby objects, closer
than $\sim$600 pc.  This is interesting because it enlarges the sample
of nearby open clusters, and also because of the increasing interest
in clusters in late stages of dissolution, as demonstrated by the
recent studies of POCRs - Possible Open Cluster Remnants - performed
by \citet{Bica2001}, and the controversial cases of \object{NGC~1252}
\citep{Baumgardt1998,Pavani2001} and \object{NGC~6994}
\citep{Odenkirchen2002,Carraro2000,Bassino2000}.

In the next section, we describe the procedures adopted in the search
for clusters and the tools used in their analysis.
Sect.~\ref{sec:res} is dedicated to the presentation of each cluster
candidate.  Finally, in
Sect.~\ref{sec:conc} we summarize the main results of this paper and
give some concluding remarks.

\section{Cluster search and analysis strategy  \label{sec:strat}}
A stellar field in which an open cluster is present, exhibits a number
of particular characteristics which reveal its existence.  The cluster
members are within a limited volume, resulting in an increased stellar
density when compared to the surrounding field.  Furthermore, they
display similar radial velocities and proper motions, which produce
clumps in phase space, and have the same age, chemical composition and
a broad range in mass, which produce distinctive sequences in a
color-magnitude diagram.  So, in theory, it is possible to distinguish
cluster members from field stars in a certain region of the sky.  One
should, however, be careful about jumping into conclusions solely
based on one of these characteristics: star counts, common proper
motions, or apparent cluster sequences in the color-magnitude diagram.
In fact, extinction patterns or chance alignment of bright stars along
the same approximate line-of-sight can give the illusion of the
presence of a cluster.  Also, certain combinations of stellar
populations and reddening can produce sequences in the color-magnitude
diagram that can be confused with a cluster sequence.

\begin{figure*}
\centering
\includegraphics[width=15cm]{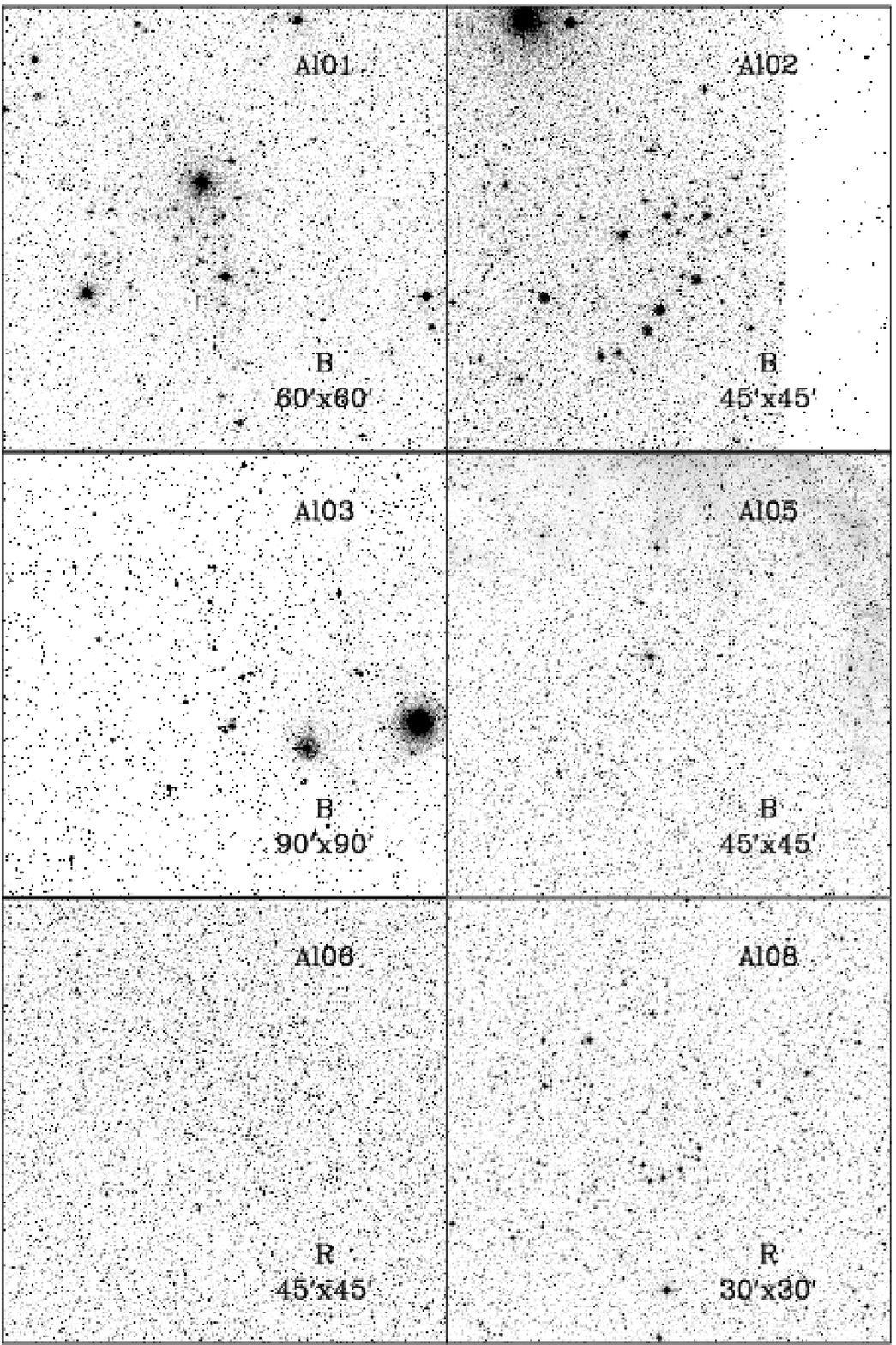}
\caption{Finding charts for Al01, 02, 03, 05, 06 and 08, taken
  from the the second generation DSS images. The field of view 
  and DSS2 band is marked in each image. North is up and East is left.}
\label{fig:charts1}
\end{figure*}

\begin{figure*}
\centering
\includegraphics[width=15cm]{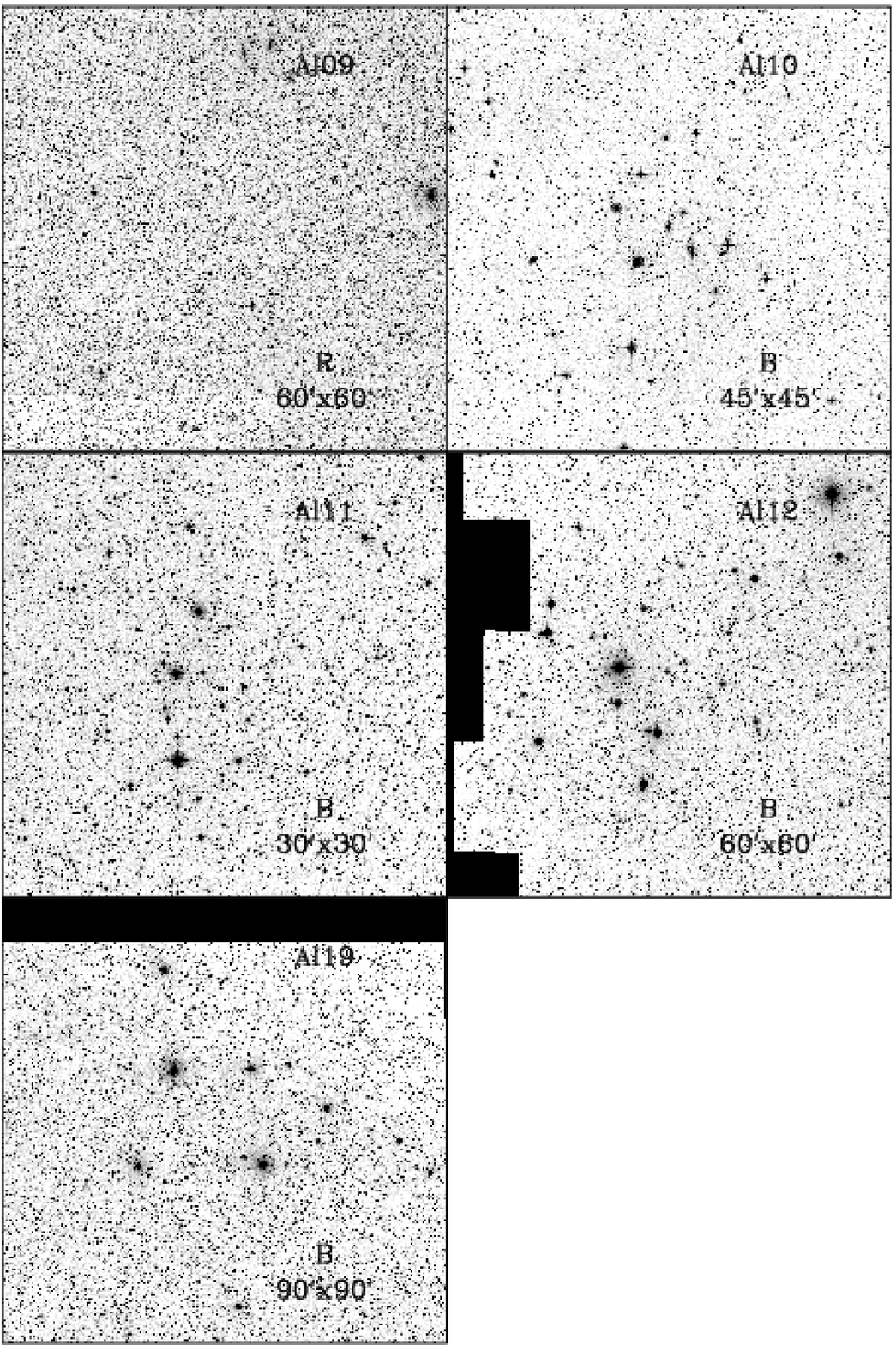}
\caption{Finding charts for Al09, 10, 11, 12, and 19, taken
  from the the second generation DSS images. The field of view 
  and DSS2 band is marked in each image. North is up and East is left.}
\label{fig:charts2}
\end{figure*}

We started by searching the Tycho2 catalog for groups of bright
stars that show simultaneously a small deviation in proper motions and
a noticeable enhancement in the local stellar density.  As a result,
several fields were selected as containing potential open cluster
candidates.  These fields were then visually inspected in the DSS
plates, and those seeming to host actual clusters were included in a
final list of open cluster candidates. The DSS images of two of the
selected fields (\object{Alessi~6} and~9) show no obvious
concentrations of stars. These two cases will be discussed below. The
finding charts for the selected objects are shown in
Figs.~\ref{fig:charts1} and~\ref{fig:charts2}.

Then, for each field, individual membership probabilities were
determined by applying the statistical method of \citet{Sanders1971}
to the Tycho2 proper motions.  For further details on the method and
its utilization we refer the reader to papers such as
\citet{Dias2001}, among others.

\begin{figure*}
\centering
\includegraphics[width=17cm]{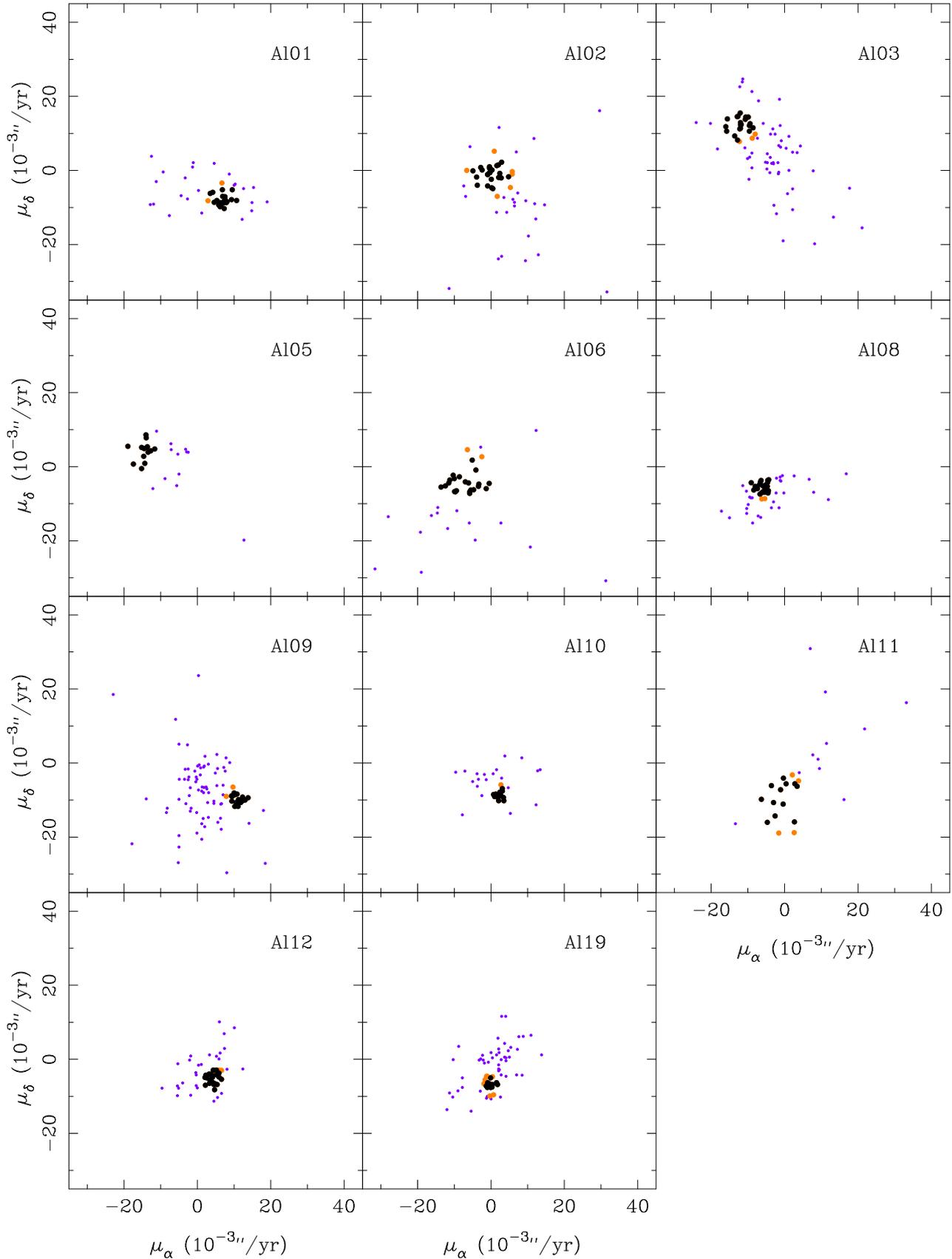}
\caption{VPDs for the 11 candidate fields.
  The big dots are stars considered members (prob. $>$ 70\% -- dark
  dots) and probable members (prob. $>$ 50\% -- light dots); non
  members (prob. $\leq$ 50\%) are represented by smaller dots.  Only
  Tycho2 stars at the 90\% completeness level ($V_{T} < 11.5$) were
  included.}
\label{fig:vpd}
\end{figure*}

For 10 of the 11 objects studied here, we found significant results
for the statistical analysis of proper motions, as well as a good
separation between the cluster and field populations in the proper
motion vector-point diagrams (hereafter VPD).  The VPDs are shown in
Fig.~\ref{fig:vpd}.

The validity of the VPD analysis was tested by comparing to a number
of control fields.  For each candidate field we took four
nearby fields (north, south, east, west) 2 cluster radii from the
center.  These comparison fields were selected with the same
diameter of the probable cluster.  For most objects we found a good
contrast between the candidate VPD and the control fields' VPDs,
further assessing that the feature observed in the candidate VPD could
correspond to an open cluster.

\begin{figure*}
\centering
\includegraphics[width=17cm]{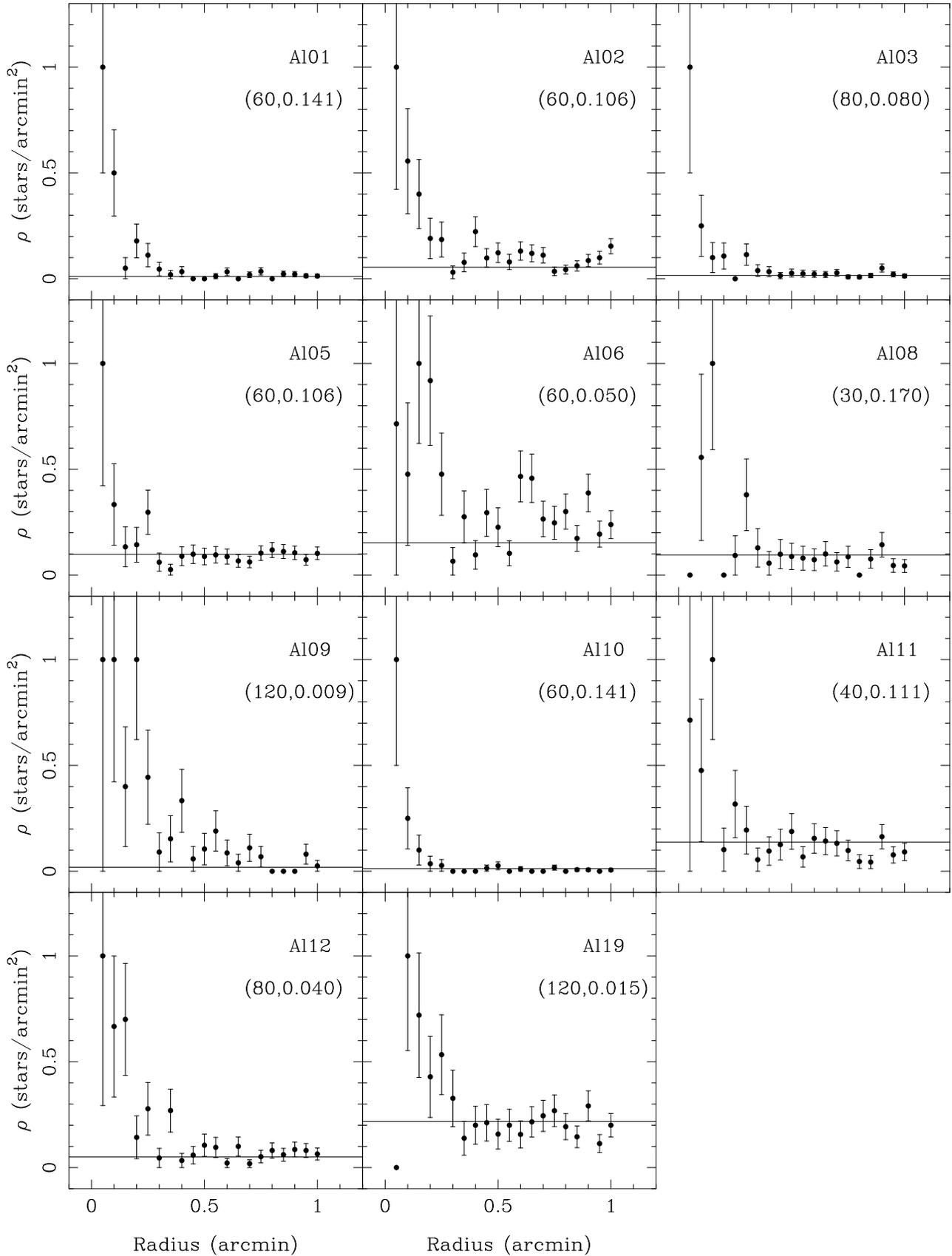}
\caption{RDPs for the 11 candidate fields.
  The numbers in brackets give the horizontal and vertical scales,
  respectively.  The vertical scale has been normalized to the highest
  density (in stars/arcmin$^2$) bin.  In the case of Al01, the data
  spans a radius of 0 to 60 arcmin, and the highest density bin has
  0.141 stars/arcmin$^2$.  Only the kinematically selected stars
  (members), defined as those with proper motions within 2$\sigma$
  from the mean value, and within the Tycho2 90\% completeness level,
  were included.}
\label{fig:rdp}
\end{figure*}

As mentioned in the beginning of this section, another indicator of
the presence of a stellar aggregate is an enhancement of the stellar
density with respect to the surrounding field.  We have plotted the
Radial Density Profile Diagrams (hereafter RDPs) shown in
Fig.~\ref{fig:rdp} to address this point.  The construction of the RDP
requires the determination of the central coordinates of each cluster
candidate. For this purpose, marginal distributions in RA and DEC were
built using Tycho2 stars with membership probability higher than
$50\%$, and brighter than $V_{T}=11.5$ mag (Tycho2 is 90$\%$ complete
in this range). The central coordinates were then taken to be the
highest peaks exhibited in each marginal distribution.

The RDPs were built using the same selection criteria (Tycho2 stars
with membership probability higher than $50\%$, and brighter than
$V_{T}=11.5$ mag).  The same selection criteria were also applied to
each of the 4 control fields in order to derive their mean densities.
Densities were calculated by counting stars in a number of concentric
rings and then dividing the counts by the area of the ring.  For each
RDP, we have also plotted a line indicating the average value of the
densities of the four nearby control fields above mentioned.  Although
we are dealing with small numbers and the statistical fluctuations are
high, we have used the proper motion criteria to isolate the potential
cluster members from the field stars.  Thus, we considered a
concentration of stars with similar motions around the candidate
cluster center as an indication of the presence of a cluster.
All the candidates display clear over-densities of Tycho2 stars.  A
further check on the first and second generation DSS images also
reveals all, but two objects (\object{Alessi~6}, 9) as slight
concentrations of bright stars, corroborating the results from the RDP
analysis.  Despite not being evident in the DSS images,
\object{Alessi~6} and~9 have been included in the final list of
candidates presented in this article. \object{Alessi~6} is an
interesting case because it could be related to the open cluster
\object{BH~164}. As for \object{Alessi~9}, all the results indicate
the presence of a genuine open cluster.

The field levels of the RDPs (horizontal line in Fig.~\ref{fig:rdp}) are
not exactly zero as would be expected if we were using only members.
This is because the RDPs were built using stars with membership
probability $> 50\%$, so some field interlopers are still present.
Each candidate's radius was estimated from the RDP as the radius where
the density profile reaches the mean density of its control fields.
Due to the limited photometric depth of the Tycho2 data, these radii
will not include the cluster coronae and should therefore be
considered as lower limits.

\begin{figure*}
\centering
\includegraphics[width=17cm]{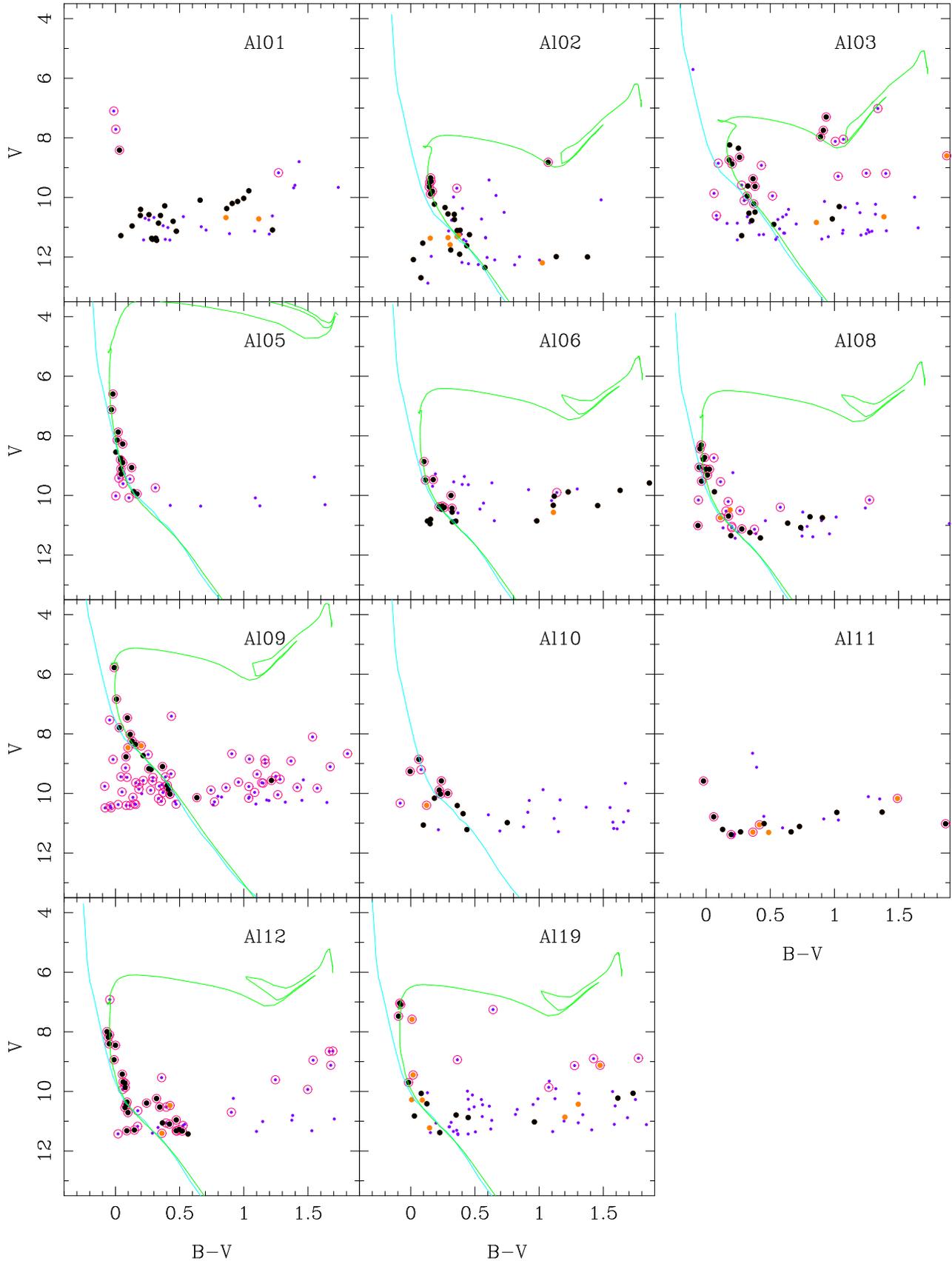}
\caption{CMDs for the 11 candidate fields. As in the VPDs,
  big dots are stars considered members (prob. $>$ 70\% -- dark dots)
  and probable members (prob. $>$ 50\% -- light dots); non members
  (prob. $\leq$ 50\%) are represented by smaller dots.  Stars with
  available spectral types are marked with circles.  The best fit ZAMS
  and isochrones have also been plotted.  Only Tycho2 stars at the
  90\% completeness level were included.}
\label{fig:cmd}
\end{figure*}

The next step was to perform a photometric analysis. Color-magnitude
diagrams (Fig.~\ref{fig:cmd}; hereafter CMDs) of the cluster fields
were plotted using the Tycho2 BV photometry transformed to the Johnson
system using the relations given in the Hipparcos catalog. Membership
probabilities were used to select  the probable cluster members,
reducing the effects of contamination by field stars and allowing the
construction of clearer CMDs.  For 7 of the 11 candidates we found
good CMDs, for 2 we have hints of a main sequence (MS), and for 2 we
do not identify any sign of a MS.  Nevertheless, although the Tycho2
magnitudes are reasonably precise for the brighter stars ($V\leq 9$),
for the fainter ones the errors in B and V become large yielding
almost useless colors for $V$ greater than $\sim 10-11$.  So our
failure to identify a MS does not necessarily reflect the nonexistence
of a cluster.

For each candidate field, there were some stars with published
spectral types.  These were used to derive intrinsic colors by
interpolating over the tables relating spectral types and colors given
by \citet{Schmidt-Kaler1982}, taking care not to use stars with
peculiar spectral features (Be, Ap, etc).  Once the intrinsic colors
were determined, individual color excesses were computed. The color
excess distributions of the member stars were then used in deriving
the mean reddenings for the candidate clusters. In a similar fashion,
the spectral types were used to estimate the absolute magnitudes, and
from there the apparent and absolute distance moduli distributions.
However, contrary to the reddening determination, where stars were
used regardless of their luminosity class, only stars of luminosity
class V (dwarfs) were used in the distance determination.  Also,
although the reddening values obtained from spectral types are quite
precise, the distances are much more uncertain.  Using the distance to
the cluster derived from spectral types as a rough estimate, we then
refined the determination via ZAMS fitting in the CMD.  The errors in
the derived reddenings, distances and ages have been estimated by
applying the sames procedures to a few well studied clusters
(\object{Blanco~1}, \object{IC~2391}, \object{Pleiades},
\object{NGC~1662} and \object{NGC~7092}). The comparison of our
values with those derived from previous detailed studies has allowed
us to estimate $\sigma_{E(B-V)} \sim 0.04$ mag, $\sigma_{dist} \sim
10\%$ and $\sigma_{age} \sim 30\%$.

Fig.~\ref{fig:cmd} shows the CMDs for the 11 fields under study.
We have superimposed the \citet{Schmidt-Kaler1982} ZAMS and solar
composition isochrones from \citet{Girardi2000iso} shifted to account
for the effects of reddening and distance. Further comments on the fits
(reddening, distance and age) will be given below in
Sect.~\ref{sec:res} when discussing each cluster.

\begin{figure*}
\centering
\includegraphics[width=17cm]{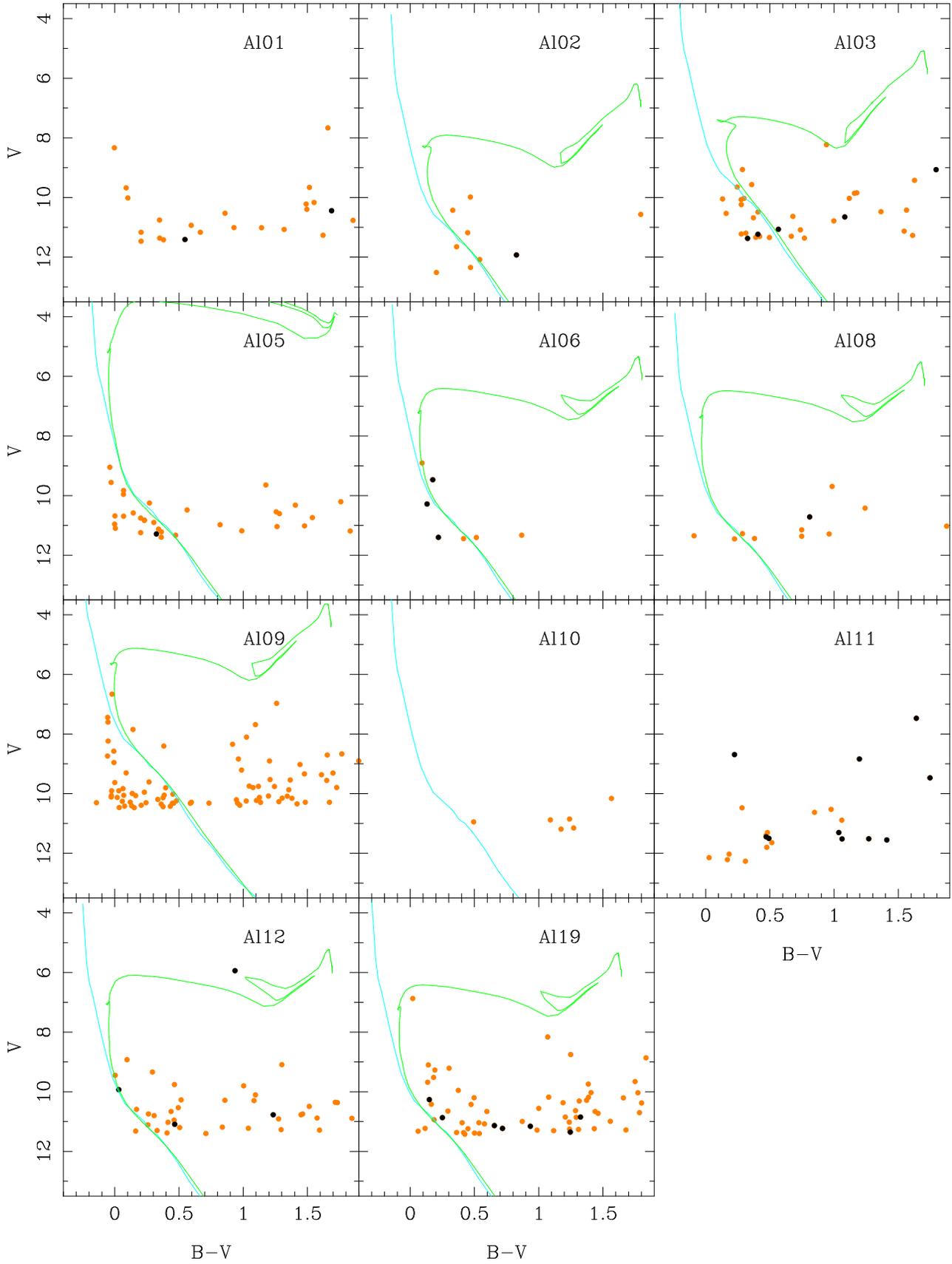}
\caption{CMDs for the combined N,S,E,W control fields. The black
  dots are stars considered members (prob. $>$ 70\%);
  all the other stars are plotted in a lighter tone.  The ZAMS and
  isochrones fitted to the cluster sequences of Fig.~\ref{fig:cmd}
  have also been plotted for comparison.  Only Tycho2 stars above the
  90\% completeness level were included.}
\label{fig:cmd_cf}
\end{figure*}

As done for the RDPs and VPDs, each candidate CMD was
  compared to 4 control fields, 2 cluster radii from the center,
  adopting the same limiting magnitude.  The combined N,S,E,W control
  fields' CMDs are shown in Fig.~\ref{fig:cmd_cf}. In order to produce
  diagrams with stars covering the same effective area as the
  clusters' CMDs and make the comparisons more meaningful, a radius of
  half a cluster radius was used for each of the N,S,E,W fields.
  Stars with P$>70$\% are plotted in black while all the others are
  plotted in a lighter tone. The ZAMS and isochrones that delineate
  the sequence detected in the clusters' CMDs have also been plotted.
  It is readily seen that the control fields do not exhibit the
  sequences seen in the clusters' CMDs. In some cases there is an
  occasional ``member'' that appears on top of, or very close to, the
  reference lines. The number of ``members'' in the control fields is
  small in all cases (they could be interlopers or even cluster stars
  that escaped), and do not define the more populated sequences seen
  in the cluster candidates' CMDs.

\section{Discussion and results \label{sec:res}}
In this section, we discuss each of the 11 investigated fields.
The main results are summarized in
Tabs.~\ref{tab:pars} and~\ref{tab:kin}.

In Tab.~\ref{tab:pars} we present the candidates' positions, angular
diameters, reddenings, distances and ages, plus the derived linear
diameters and velocity dispersions.  The adopted procedures were
described in Sect.~\ref{sec:strat} and the results confirm that if
real, the clusters are likely to be situated in the solar vicinity.
The results from the kinematic analysis are given in
Tab.~\ref{tab:kin}.  Finally, in Tab.~\ref{tab:hip} we list the
Hipparcos members found in the regions of several candidates.  The
identification of possible Hipparcos members allows us to estimate the
distance to these objects in a different way.  Because of the small
number of such stars, the mean cluster parallax was simply taken as
the mean of the individual stellar parallaxes, and no special
corrections were applied. When only one measurement was available, it
was adopted as the cluster parallax. The distances obtained from the
CMDs are also listed for comparison purposes.  Some clusterings
(\object{Alessi~2}, 6, 12, 19) had Hipparcos members in their fields
with large errors due to their small (or even negative) parallaxes.
These stars are also listed in Tab.~\ref{tab:hip} for the sake of
completeness. Overall, despite that the small number and precision of
the parallaxes of member stars does not allow rigorous distance
determinations, they do provide a very useful consistency check of the
photometrically derived distances.

Considering together the results obtained from each point of view -
astrometric, photometric and star counts - nearly all the investigated
fields show signs of the existence of open clusters. Except
  for 2 fields (\object{Alessi~1} and 11) all the candidates show
  signs of a cluster sequence in the CMD, while the comparison fields
  show nothing.  A final conclusion regarding the true nature and
parameters of the clusterings will require, however, more precise and
deeper photometry.

We will now proceed to discuss each of the 11 investigated fields.

\begin{table*}[ht]
\caption[]{Basic parameters of the investigated cluster candidates. 
The equatorial coordinates are given in 
J2000.0. 
The cluster diameter, {\em D}, is  given in both arcsec and parsec;
{\em d} is the distance to the cluster;
$\log t$ is the logarithm of the age (in years); {\em Disp} is the velocity
dispersion within the cluster.
}\label{tab:pars}
\begin{center}
\begin{tabular}{lccccccccccc}
\hline \hline
Name &    $l$   &  $b$ & $\alpha(2000)$ & 
$\delta(2000)$ & $D$ 
&        d & $(m-M)_0$ & $E(B-V)$ & $\log t$ & D & Disp\\
&&&&& (${}^{\prime}$)
       & (pc) &  &  &  & (pc) & ($km/s$) \\

\hline
\object{Alessi~1}  & $123\fdg 2$ & $-13\fdg 3$ & $00^{\rm h}53^{\rm m}00.0^{\rm s}$ & $+49\degr 33\farcm 0$ & 54 
 &     &      &      &      &                &     \\
\object{Alessi~2}  & $152\fdg 3$ & $+06\fdg 4$ & $04^{\rm h}46^{\rm m}24.0^{\rm s}$ & $+55\degr 15\farcm 0$ & 36  
 & 500 & 8.50 & 0.18 & 8.50 & \phantom{1}5.2 & 6.6 \\
\object{Alessi~3}  & $257\fdg 8$ & $-15\fdg 4$ & $07^{\rm h}16^{\rm m}24.0^{\rm s}$ & $-46\degr 36\farcm 0$ & 72  
 & 290 & 7.30 & 0.11 & 8.70 & \phantom{1}6.1 & 2.9 \\
\object{Alessi~5}  & $288\fdg 1$ & $-02\fdg 1$ & $10^{\rm h}42^{\rm m}54.0^{\rm s}$ & $-61\degr 12\farcm 0$ & 36  
 & 400 & 8.00 & 0.15 & 7.60 & \phantom{1}4.2 & 4.2 \\
\object{Alessi~6}  & $313\fdg 6$ & $-05\fdg 6$ & $14^{\rm h}40^{\rm m}12.0^{\rm s}$ & $-66\degr 10\farcm 5$ & 36  
 & 440 & 8.20 & 0.19 & 8.20 & \phantom{1}4.6 & 7.0 \\
\object{Alessi~8}  & $326\fdg 5$ & $+04\fdg 3$ & $15^{\rm h}29^{\rm m}30.0^{\rm s}$ & $-51\degr 12\farcm 5$ & 24  
 & 570 & 8.80 & 0.09 & 8.15 & \phantom{1}4.0 & 4.6 \\
\object{Alessi~9}  & $344\fdg 3$ & $-09\fdg 1$ & $17^{\rm h}44^{\rm m}00.0^{\rm s}$ & $-47\degr 00\farcm 0$ & 144  
 & 190 & 6.45 & 0.07 & 8.40 & \phantom{1}8.0 & 1.1 \\
\object{Alessi~10} & $31\fdg 6$ & $-21\fdg 0$ & $20^{\rm h}04^{\rm m}48.0^{\rm s}$ & $-10\degr 31\farcm 5$ & 36  
 & 380 & 7.90 & 0.18 &      & \phantom{1}4.0 & 2.0 \\
\object{Alessi~11} & $59\fdg 9$ & $-10\fdg 3$ & $20^{\rm h}21^{\rm m}12.0^{\rm s}$ & $+18\degr 24\farcm 0$ & 28  
 &     &      &      &      &                &               \\
\object{Alessi~12} & $67\fdg 4$ & $-11\fdg 4$ & $20^{\rm h}43^{\rm m}18.6^{\rm s}$ & $+23\degr 46\farcm 5$ & 64  
 & 540 & 8.65 & 0.08 & 8.10 & 10.1            & 3.5 \\
\object{Alessi~19} & $40\fdg 1$ & $+12\fdg 7$ & $18^{\rm h}18^{\rm m}24.0^{\rm s}$ & $+12\degr 01\farcm 5$ & 84  
 & 550 & 8.70 & 0.03 & 8.20 & 13.4           & 4.1 \\
\hline
\end{tabular}
\end{center}
\end{table*}

\begin{table*}[ht]
\caption[]{Results from the proper motion analysis. 
The meaning of the
symbols are as follows: 
$N_{c}$ is the number of cluster stars;
$N_{f}$ is the number of field stars;
$\mu_{\alpha}cos{\delta}$ and $\mu_{\delta}$ are the  proper
motion components in mas/yr;
$\sigma$ is the dispersion of cluster stars' proper motions;
$\sigma\mu_{\alpha}cos{\delta}$ and $\sigma \mu_{\delta}$  are the
dispersions of the components of the field stars' proper motions; $\theta$
is the orientation angle of the minor axis of elliptical field star
proper motion distribution.}\label{tab:kin}
\begin{center}
\begin{tabular}{lcccc|cccccc}
\hline \hline
&\multicolumn{4}{c|}{Cluster} & \multicolumn{6}{c}{Field}\\
Region    & 
$\mu_{\alpha}cos{\delta}$ & 
$\mu_{\delta}$ & 
$\sigma$ &
$N_{c}$  & 
$\mu_{\alpha}cos{\delta}$ & 
$ \mu_{\delta}$ & 
$\sigma\mu_{\alpha}cos{\delta}$ &  
$\sigma \mu_{\delta}$ & 
$N_{f}$ & 
$\theta$ \\  
\hline
\object{Alessi~1}  & \phantom{-1}6.7 & \phantom{1}-7.8 & 1.7 & 28 & \phantom{-}2.2 & \phantom{1}-5.3 & 10.3           & \phantom{1}5.8 &  19 & 278.0           \\ 
\object{Alessi~2}  & \phantom{-1}0.2 & \phantom{1}-1.1 & 2.8 & 23 & \phantom{-}6.2 & \phantom{-}-9.5 & 11.7           & 14.6           &  33 & \phantom{1}77.9 \\ 
\object{Alessi~3}  & -11.5           & \phantom{-}12.4 & 2.1 & 22 & -3.8           & \phantom{-1}4.4 & 12.7           & 11.6           &  55 & \phantom{1}35.2 \\ 
\object{Alessi~5}  & -14.3           & \phantom{-1}4.4 & 2.2 & 13 & -5.4           & \phantom{-1}0.1 & \phantom{1}7.8 & \phantom{1}9.8 &  13 & 335.1           \\ 
\object{Alessi~6}  & \phantom{1}-7.3 & \phantom{1}-4.3 & 3.4 & 18 & -9.2           & -14.1           & 19.8           & 17.0           &  23 & 244.5           \\ 
\object{Alessi~8}  & \phantom{1}-6.1 & \phantom{1}-5.8 & 1.7 & 22 & -3.8           & \phantom{1}-7.8 & \phantom{1}8.5 & \phantom{1}5.0 &  28 & 255.7           \\ 
\object{Alessi~9}  & \phantom{-}10.9 & \phantom{1}-9.7 & 1.3 & 21 & \phantom{-}1.0 & \phantom{1}-9.8 & 10.5           & 13.4           &  78 & \phantom{1}74.6 \\ 
\object{Alessi~10} & \phantom{-1}2.3 & \phantom{1}-8.6 & 1.1 & 17 & \phantom{-}0.9 & \phantom{1}-4.8 & \phantom{1}6.7 & \phantom{1}4.4 &  16 & 268.0           \\ 
\object{Alessi~11} & \phantom{1}-0.3 & \phantom{1}-9.6 & 4.4 & 12 & \phantom{-}8.4 & \phantom{-1}1.5 & 16.2           & 16.6           &  15 & 133.1           \\ 
\object{Alessi~12} & \phantom{-1}4.3 & \phantom{1}-5.3 & 1.4 & 28 & \phantom{-}2.0 & \phantom{1}-2.9 & \phantom{1}7.1 & \phantom{1}6.2 &  28 & 161.8           \\ 
\object{Alessi~19} & \phantom{-1}0.4 & \phantom{-}-6.6 & 1.6 & 18 & \phantom{-}1.2 & \phantom{1}-2.5 & \phantom{1}8.2 & \phantom{1}7.3 &  52 & 147.5           \\ 
\hline
\end{tabular}
\end{center}
\end{table*}

\begin{table*}[ht]
\caption[]{Hipparcos stars in the region of some investigated cluster
 candidates.
The individual parallaxes and their errors are given, as well as the
membership probability ({\em P} in $\%$) . {\em Par} is the adopted cluster
parallax, {\em d} is the distance derived
from the adopted parallax, and {\em d(CMD)} is the distance derived
from the CMD. Stars marked with a colon (:) were not considered in the
computation of the parallax.}\label{tab:hip}
\begin{center}
\begin{tabular}{lr@{}lrccccc}
\hline\hline
Name  &  \multicolumn{2}{c}{HIP} &  Plx  & ePlx  &    P  & Par           &  d    & d(CMD) \\
\hline
\object{Alessi~2}  &  22283 &    & -0.54 &  1.67 &    94 &               &       & 500  \\
\object{Alessi~3}  & :35207 & AB &  2.79 &  1.02 &    85 & 4.17          & 240   & 290  \\
                   &  34974 &    &  4.17 &  0.59 &    88 &               &       &      \\
\object{Alessi~5}  & :52436 & AC &  3.86 &  1.08 &    95 & 3.11          & 320   & 400  \\
                   & :52437 &  B &  5.89 &  3.59 &    83 &               &       &      \\
                   &  52534 &    &  3.11 &  0.84 &    96 &               &       &      \\
\object{Alessi~6}  &  71613 &    &  0.19 &  1.20 &    92 &               &       & 440  \\
                   &  71560 &    & -1.34 &  1.91 &    92 &               &       &      \\
                   &  71674 &    & -1.38 &  1.62 &    89 &               &       &      \\
                   &  71435 &    & -2.05 &  1.29 &    91 &               &       &      \\
\object{Alessi~9}  &  86552 &    &  4.94 &  0.90 &    95 & 4.94          & 200   & 190  \\
                   & :86658 & AB &  5.59 &  2.62 &    84 &               &       &      \\
\object{Alessi~10} &  98899 &    &  2.72 &  1.39 &    97 & 2.82          & 355   & 380  \\
                   &  98904 &    &  2.92 &  1.71 &    39 &               &       &      \\
\object{Alessi~12} & 102277 &    &  0.62 &  0.89 &    92 &               &       & 540  \\
                   & 102374 &    &  0.22 &  0.84 &    90 &               &       &      \\
                   & 102287 &    &  0.95 &  0.91 &    91 &               &       &      \\
\object{Alessi~19} &  89767 &    &  0.50 &  0.78 &    81 &               &       & 550  \\
                   &  89665 &    &  1.56 &  0.83 &    76 &               &       &      \\
                   &  89798 &    &  1.03 &  0.88 &    67 &               &       &      \\
                   &  89776 &    &  1.07 &  0.96 &    81 &               &       &      \\
\hline
\end{tabular}
\end{center}
\end{table*}

\subsection{\object{Alessi~1}}
This candidate is a large ($54^{\prime}$), isolated group of stars
located relatively far from the Galactic plane ($b=-13\fdg 3$).  It
stands fairly well in the DSS image (Fig.~\ref{fig:charts1}), and
shows significant features in both the RDP and VPD diagrams. All this
evidence points to the presence of an open cluster. In fact, the
kinematic analysis yielded 28 probable members.

The photometric result is not good since the CMD presented in
Fig.~\ref{fig:cmd} does not exhibit any  cluster sequence. Instead,
we find one bright member followed by a gap of $\sim 1.5$ mag and
then number of fainter stars badly spread out in the CMD. These
fainter stars have $B$ or $V$ fainter than $\sim$ 10.5 mag, so the
scatter is expected to be mainly a result of the photometric precision
of the Tycho2 data.  
A final confirmation of the reality of this cluster and the
determination of its fundamental parameters will require
deeper and more precise photometric data.

\subsection{\object{Alessi~2}}
\object{Alessi~2} resembles a slightly fainter and smaller version of
\object{Alessi~1} in the DSS image (Fig.~\ref{fig:charts1}). The VPD
(Fig.~\ref{fig:vpd}) shows a noticeable concentration centered nearly
at the 0,0 point. The RDP (Fig.~\ref{fig:rdp}) shows a significant
over-density in the cluster core when compared to the control fields.
Despite its faintness, the CMD (Fig.~\ref{fig:cmd}) is fairly well
defined with a visible main-sequence, turn-off and even one star
considered as member in the giant branch.

The only Hipparcos member in the field happens to have a negative
parallax (see Tab.~\ref{tab:hip}).  We note, however, that the
distance estimated from the CMD is 500 pc, for which the errors in the
Hipparcos parallaxes are greater than 100\% (distance greater than
$\sim$ 400 pc) and therefore negative parallaxes are expected. In this
sense, although useless in providing an independent distance estimate,
the small (in this case, negative) parallax is consistent with the CMD
derived distance.

\subsection{\object{Alessi~3}}
\object{Alessi~3} stands out as a chain of bright stars that dominates
the entire 90 arcmin of the DSS image (Fig.~\ref{fig:charts1}).  As
\object{Alessi~1}, this candidate lies far from the plane of Milkyway
with $Z\sim -78$ pc.  The region has a nice VPD (Fig.~\ref{fig:vpd})
with a clear separation between cluster members and field stars.  The
RDP (Fig.~\ref{fig:rdp}) also agrees well with the existence of a
concentration of kinematically related stars.  Although somewhat
scattered, the CMD (Fig.~\ref{fig:cmd}) shows the characteristic
features expected for a cluster: a main sequence, a turn-off and three
stars considered members in the giant branch.  The photometric
analysis yields \object{Alessi~3} as the oldest cluster candidate in
the sample with probably half a billion years.

The field contains two Hipparcos stars considered as members but 
with very different parallaxes (see Tab.~\ref{tab:hip}). 
One of them, 
\object{HIP35207AB}, belongs to a multiple system so we expect
its parallax not to be so reliable. Based on the other star,
\object{HIP34974}, we find a distance of 240 pc which is in excellent
agreement with the 290 pc derived photometrically.

A subsequent verification in Tycho2 shows probable members out
to a diameter of 80$^{\prime}$ or even more, including two other
Hipparcos (giant) stars which are also possible members.
This is not surprising since this candidate is presumably a
nearby one.

\subsection{\object{Alessi~5}}
Apparently, this is the most compact candidate in the
investigated sample.  It appears in the DSS plates
(Fig.~\ref{fig:charts1}) as a few bright
stars concentrated around the bright multiple star \object{DUN~97}
(\object{HIP~52437}) and seen in projection against a crowded
  area of the Carina arm.  The concentration of stars is
easily seen as the sharp central peak displayed by its RDP
(Fig.~\ref{fig:rdp}). Its VPD (Fig.~\ref{fig:vpd}) shows an evident,
although poor, concentration of members.

The CMD (Fig.~\ref{fig:cmd}) shows a clear main sequence.  The absence
of stars in the giant branch and a nearly vertical main sequence
containing mainly B3-B8 stars are indicative of a young aggregate.
As with many young open clusters, the absence of clear evolutionary
features together with poor statistics result in somewhat uncertain
age determinations.  There are three Hipparcos members
(Tab.~\ref{tab:hip}) in the \object{Alessi~5} region and the distance
estimated from their parallaxes (320 pc) is consistent with the
photometric distance (400 pc).  We do note, however, that two of the
Hipparcos stars are multiple systems that present large errors in
their parallaxes.  

It is interesting to point out that the nearby classical cepheid
\object{EY~Car} (\object{HIP~52380}) is certainly not a member of
\object{Alessi~5} despite the coincidence of its kinematic parameters
in the Hipparcos catalog ({$\mu_{\alpha}cos{\delta} = -9.12$} mas/yr
and {$\mu_{\delta}= +4.28$} mas/yr).  According to the distance
determined by \citet{Berdnikov2000} (2.02 kpc) this cepheid is located
behind of the candidate, probably in the Carina arm.

\subsection{\object{Alessi~6}}
The DSS image of the field of \object{Alessi~6}
(Fig.~\ref{fig:charts1}) does not reveal any evident clustering of
stars. Interestingly, the VPD analysis allowed to identify a clump of
stars with high membership probabilities whose RDP
(Fig.~\ref{fig:rdp}) reveals a significant spatial concentration of
stars with similar kinematics.  Although the concentration in the VPD
(Fig.~\ref{fig:vpd}) is not very tight, the statistical solution is
compatible with the existence of an open cluster, and there is a good
separation between the distribution of the stars considered as members
and the field population.

The CMD (Fig.~\ref{fig:cmd}) shows a poorly populated cluster sequence
with no clear evolutionary features, which makes the derived age quite
uncertain.  As with most of the fields presented in this work, the
scatter in the lower MS is likely to be due mainly to photometric
errors, although the redder stars could be field giant interlopers.

Similarly to \object{Alessi~2}, the Hipparcos members in the field
have large errors, being most of them negative. Again, no useful
distance can be derived from the parallaxes although they do support
the indication given by the CMD that \object{Alessi~6} is beyond
$\sim$ 400 pc.

The cluster candidate \object{Alessi~6} is an interesting case because
it forms a close pair with the nearby cluster \object{BH~164} (the
centers of both aggregates are 51$^{\prime}$ apart).  Curiously, the
\citet{Lynga1987} coordinates of \object{BH~164}, taken from
\citet{vandenBergh1975}, show no evidence of a cluster.  New
coordinates, fundamental parameters and mean absolute proper motions
of \object{BH~164} have been recently determined by
\citet{Moitinho2003a} and by \citet{Kharchenko2003}. Both studies
were based on Tycho2 data.  

Besides their small separation, both clusters have similar ages and
are located at 440 pc \citep{Moitinho2003a} which corresponds to a
separation of about 15 pc. Their close distance could indicate a real
physical relation between both groups of stars.  On the other hand,
\citet{Kharchenko2003} puts \object{BH~164} at a distance of 540 pc,
which would imply that it is not physically related to
\object{Alessi~6}.  Despite the different distance determinations,
both studies give similar proper motions for \object{BH~164}
($\mu_{\alpha}cos{\delta}=-7.1 \pm 0.4$ mas/yr; $\mu_{\delta}=-10.9
\pm 0.4$ mas/yr), which are not very different from those estimated
for \object{Alessi~6} ($\mu_{\alpha}cos{\delta}=-7.3 \pm 3.4$ mas/yr;
$\mu_{\delta}=-4.3 \pm 3.4$ mas/yr).  A final word regarding the true
nature of the possible pair \object{BH~164}/\object{Alessi~6} will
require deeper photometric data and radial velocities for both
clusters.

\subsection{\object{Alessi~8}}
This is a possible poor cluster in the Lupus-Norma border.
\object{Alessi~8} is seen in the DSS image (Fig.~\ref{fig:charts1}) as
a compact arc of a few bright stars surrounded by fainter stars almost
not detached from the background.

The VPD (Fig.~\ref{fig:vpd}) shows an evident clump of stars which are
spatially concentrated, as expressed by the RDP (Fig.~\ref{fig:rdp}).
The CMD exhibits a fairly defined sequence with a 1 mag. gap between
$V\sim 10 - 11$.  All these results support the existence of a cluster
in the field.

\subsection{\object{Alessi~9}}
Despite not being easily seen in the DSS image
(Fig.~\ref{fig:charts2}), \object{Alessi~9} is from all points of view
(astrometric, photometric and star counts) one of the best candidates
studied here.  The RDP (Fig.~\ref{fig:rdp}) shows an obvious
concentration of stars and the VPD (Fig.~\ref{fig:vpd}) exhibits a
clear segregation of the cluster and field populations.  Due to the
aggregate's large extent (about $140^{\prime}$) and richness of the
\object{Milkyway} background, only Tycho2 stars brighter than $V_{T} <
10.5$ where considered in the astrometric analysis in order to reduce
the contamination by field stars.  The CMD (Fig.~\ref{fig:cmd}) has a
well defined morphology (ZAMS and turn-off) that indicate the
existence of a true open cluster and allow to obtain good
reddening, distance and age estimates.  We find that \object{Alessi~9}
is the nearest (190 pc) open cluster in the sample.

Two interesting Hipparcos stars were found in the region of the
cluster (see Tab.~\ref{tab:hip}): one analyzed member
(\object{HIP~86552}) and one possible member (\object{HIP~86658}) that
is a $\beta$Lyr type eclipsing binary. Comparison of the distance
obtained from the CMD analysis and that provided by the Hipparcos
parallaxes shows a good agreement between both values.

\subsection{\object{Alessi~10}}
This is a very isolated group of late $B$ and early $A$ stars situated
far from the Galactic plane ($Z\sim -140$ pc), in a poor field with
almost nothing but faint anonymous galaxies.  It stands as a chain of
bright stars in the DSS image (Fig.~\ref{fig:charts2}).  This group is
very interesting because its VPD (Fig.~\ref{fig:vpd}) shows one of the
best separations between field and cluster populations in the sample
of candidates presented here.  Its RDP (Fig.~\ref{fig:rdp}) shows a
clear density enhancement of kinematically related stars relative to
its very underpopulated background. The CMD shown in
Fig.~\ref{fig:cmd} is not very clear, exhibiting considerable
spreading, which results in rather uncertain estimates of both
reddening and distance and no age estimate.  Despite the limited
quality of the data, analysis of the published spectral types for the
8 brightest members seems to indicate that the scatter could be due to
variable reddening.  A more conclusive analysis will require more and
deeper astrometric and photometric data. Nevertheless, the existence
of a physical cluster seems to be the best explanation for the presence
of a tight group of kinematically similar early type stars far from
the Galactic plane.

The field of \object{Alessi~10} also hosts two Hipparcos stars
(\object{HIP~98899} and \object{HIP~98904}) with high membership
probabilities. Once again we find a good agreement between the
distance obtained from the Hipparcos parallaxes of these two stars and
that derived from the photometric data and spectral types.  Although
the number of detected members is small, we do find a few possible
members inside an area of nearly one degree, almost twice the
cluster's estimated diameter.  

\subsection{\object{Alessi~11}}
This group lies in an interesting field with three relatively bright
variable stars in its radius and a fourth one in the immediate
vicinity:
\object{BX~Del} a pop II cepheid; 
\object{BW~Del} an eclipsing algol-type binary;
\object{UV~Del} a semi-regular and 
\object{GR~Sge} an irregular variable. 
The available studies of these stars do not allow us to assess the
possibility of their association to \object{Alessi~11}.  According to
the only information found in the literature, \object{BX~Del} is at a
distance of 1.355 kpc \citep{Beers1995}.

The analyzed group of stars is not well detached from the background as
its RDP indicates (Fig.~\ref{fig:rdp}).
The VPD shows a large scatter which is expressed in the standard
deviation obtained by the statistical analysis (Tab.~\ref{tab:kin}).
Even more dramatically than in the case of \object{Alessi~1},
the faint magnitudes of the Tycho2 stars render
a useless CMD for \object{Alessi~11}, allowing no estimates
of reddening, distance or age. 

Due to the limitations of the present dataset, 
the nature of \object{Alessi~11} remains quite doubtful, although the
RDP seems to support the existence of a cluster, and so does the kinematic
analysis (at least marginally).

\subsection{\object{Alessi~12}}
\object{Alessi~12} is a large ($64^\prime$) sparse group of the stars,
as seen in the DSS image (Fig.~\ref{fig:charts2}).  Nevertheless, its
RDP (Fig.~\ref{fig:rdp}) shows an evident concentration of
kinematically related stars well above the background. The VPD
(Fig.~\ref{fig:vpd}) also displays a good separation between the
cluster and field populations.  The CMD (Fig.~\ref{fig:cmd}) displays
a well defined main sequence and turn-off which allow us to be fairly
confident on the estimated reddening, distance and age.

The field of \object{Alessi~12} contains three Hipparcos proper motion
members (see Tab~\ref{tab:hip}).  Again, as with \object{Alessi~2}
and~6, the errors in the parallaxes are higher than 100\%, which we take
as indicative of a distance greater than $\sim 400$ pc.

\subsection{\object{Alessi~19}}
This is another very large group of bright stars. Its RDP reveals an
evident concentration of kinematically related stars and the VPD shows
a very small dispersion and good separation between members and
non-members.  The CMD, however, is not very clear since it is poor and
exhibits a large gap that results in a not well defined ZAMS and
turn-off. The large derived diameter of 14 pc is not characteristic of
a standard open cluster and may indicate that \object{Alessi~19} could
be possibly a loose association.

The 4 Hipparcos members presented in Tab.~\ref{tab:hip} have errors
close to 100\% in their parallaxes and do not allow a usefull distance
estimate. Once more, as in the cases of \object{Alessi~2}, 6, and~12,
we take this as indicative of a possible stellar aggregate
located at a distance greater than $\sim 400$, which is consistent
with the 550 pc derived from the CMD.

\section{Summary \label{sec:conc}}

The Sanders technique applied to the Tycho2 proper motions has proved
to be a powerful tool to search for new, bright and nearby open cluster
candidates.  

Radial density profiles, built using stars sharing similar proper
motions, have revealed the existence of 11 previously undetected star
clusterings. Because the stars in each cluster candidate have similar
kinematics, their spatial concentration is not likely due to
projection or extinction effects.  In order to verify whether these
concentrations correspond to real open clusters, CMDs were built using
Tycho2 photometry.  Except for two objects (\object{Alessi~1} and~11),
we have been able to identify cluster sequences, further assessing the
reality of the stellar aggregates.  The kinematic analysis yielded the
candidates' mean proper motions and membership probabilities.  The
CMDs together with published spectral types have allowed us to obtain
preliminary estimates of their reddenings, distances and ages.  We
found that all the studied objects have diameters larger than
$20^\prime$ and are closer than 0.6 kpc, which puts them in the solar
vicinity.  They are also relatively young or of intermediate age.
  
Regarding the nature of the open cluster candidates, we find that
Alessi~2, 3, 5, 6, 8, 9, 10 and 12 are likely open clusters.
Alessi~19 is a possible loose cluster or association and finally,
Alessi~1 and 11 are ambiguous cases that need more and deeper data
before reaching any further conclusion.

The characteristics of most of the studied clusters (morphology, poor
core with the presence of remote members, among others) could indicate
that some of them are dissolving clusters. We expect that more data
(deeper imaging and radial velocities) will allow to establish if some
of the clusters are indeed OCRs, and if so, whether they have been
dissolved by natural dynamical evolution, or by striping due to
encounters with giant interstellar clouds or tidal interactions with
the Galactic disk.

\begin{acknowledgements}
  The authors wish to thank J.R.D. L\'epine for reading and commenting
  the manuscript, and the anonymous referee whose
  comments helped making the text clearer and more precise.
  Extensive use has been made of the SIMBAD, WEBDA and DSS databases
  and of the {\em Sky Charts (http://astrosurf.com/astropc/)}
  facility.  AM acknowledges financial support by CONACyT (Mexico;
  project I33940-E), DGAPA (Mexico; project IN111500), and FCT
  (Portugal; grant BPD/20193/99).  WSD is supported by FAPESP (grant
  number 99/11781-4).
\end{acknowledgements}


\end{document}